\documentclass{PoS}
\usepackage{amsmath}
\usepackage{amssymb}
\usepackage{mathrsfs}
\usepackage{setspace}
\usepackage{wrapfig}
\usepackage{color}
\allowdisplaybreaks[1]
\usepackage{graphicx}
\usepackage{grffile}
\usepackage{subfigure}

\def\slashchar#1{\setbox0=\hbox{$#1$} 
\dimen0=\wd0 
\setbox1=\hbox{/} \dimen1=\wd1 
\ifdim\dimen0>\dimen1 
\rlap{\hbox to \dimen0{\hfil/\hfil}} 
#1 
\else 
\rlap{\hbox to \dimen1{\hfil$#1$\hfil}} 
/ 
\fi}

\newcommand{\td}{{\rm{d}}}

\title{Towards non-perturbative matching of three/four-flavor Wilson coefficients with a position-space procedure}

\ShortTitle{NP matching of 3/4-flavor Wilson coefficients in position-space}

\author{\speaker{Masaaki Tomii}\\
        Physics Department, Columbia University, New York 10027, USA\\
        E-mail: \email{mt3164@columbia.edu}}

\abstract{
We propose a strategy to non-perturbatively match the Wilson coefficients
in the three- and four-flavor theories, which uses two-point Green's functions
of the corresponding four-quark operators at long distances.
The idea is refined by combining with the spherical averaging technique, which
enables us to convert two-point functions calculated on the lattice into
continuous functions of the distance $|x-y|$ between two operators.
We also show the result for an exploratory calculation of two-point functions
of the $\Delta S=1$ operators $Q_7$ and $Q_8$ that are in the $(8_L,8_R)$
representation of ${\rm SU(3)}_L\times{\rm SU(3)}_R$ 
and mix with each other.
}

\FullConference{The 36th Annual International Symposium on Lattice Field Theory - LATTICE2018\\
		22-28 July, 2018\\
		Michigan State University, East Lansing, Michigan, USA.}

\begin{document}

\section{Introduction}
\label{sec:intro}

Lattice calculations of weak matrix elements play an important role in searching 
for physics beyond the Standard Model. 
Weak-boson exchanges in low-energy processes can be reduced to an effective
weak Hamiltonian composed of four-quark operators by integrating out
the weak bosons and quarks heavier than the renormalization scale $\mu$.
Then, the information at high energies $> \mu$ is expressed in terms of
the Wilson coefficients, the coefficients of the four-quark operators in the
weak Hamiltonian.

For many processes, the corresponding Wilson coefficients are
known to one- or two-loop level in perturbative QCD both in the
$\rm\overline{MS}$ and RI/(S)MOM schemes.
Therefore the four-quark operators need to be renormalized at the
same renormalization scale and in the same scheme as the Wilson
coefficients to construct the proper weak Hamiltonian.
The RI/(S)MOM scheme is more straightforward than the $\rm\overline{MS}$
scheme for actual lattice calculations. 

We also need to match the number of flavors in the renormalization
scheme of the Wilson coefficients and the four-quark operators.
The perturbative calculation of the Wilson coefficients in the
three-flavor theory needs a conversion from those in the four-flavor
theory at an energy scale below the charm threshold $m_c \simeq 1.3$~GeV,
where perturbative calculation is quite ambiguous.
While the difference between the three- and four-flavor Wilson coefficients
is not significant if the form of the four-quark operators in the three- and
four-flavor theories is the same and the sea charm effect is not significant,
the issue is more serious when the charm quark can be involved in the
four-quark operators in the four-flavor theory.
In such a case, it is preferable to introduce the four-quark operators in
the four- or five-flavor theory so that we do not need the three-flavor
Wilson coefficients. 
However, if the lattice ensemble on which matrix elements are calculated
is too coarse $a^{-1} \le 2$~GeV to introduce the charm quark, 
a non-perturbative matching of the Wilson coefficients between the three-
and four-flavor theories is needed.
The RBC and UKQCD collaborations are facing this issue in the
calculation of direct CP-violating effects in $K\to\pi\pi$ decays.
Their original result contained 12\% systematic uncertainty
because of the perturbative matching of the Wilson coefficients
Ref.~\cite{Bai:2015nea}.

In this work, we formulate a strategy to nonperturbatively match the
three- and four-flavor Wilson coefficients and perform some exploratory
calculations.
As explained in Section~\ref{sec:matching},
the strategy basically uses the two-point functions of four-quark operators,
which are gauge invariant and prevent mixing with gauge-noninvariant
operators and operators that are forbidden by equations of motion.
In order to take the continuum limit of the matching matrix accurately, we
propose to take the spherical average of two-point functions \cite{Tomii:2018zix},
which is briefly explained in Section~\ref{sec:spheave}.
Some exploratory results for the spherical average of two-point functions of the
three-flavor operators in the $\rm(8_L,8_R)$ representation are shown in
Section~\ref{sec:4q2pt}.

\section{Non-perturbative three/four-flavor matching of Wilson coefficients}
\label{sec:matching}

We start with the weak Hamiltonian 
\begin{equation}
H_W = \sum_iw_{n_f,i}^{S_{n_f'}}(\mu) O_{n_f,i}^{S_{n_f'}}(\mu)\Big|_{n_f'=n_f}
= {w_{n_f}^{S_{n_f'}}(\mu)}^TO_{n_f}^{S_{n_f'}}(\mu)\Big|_{n_f'=n_f},
\label{eq:HW}
\end{equation}
where $\mu$ denotes the renormalization scale in a scheme
indicated by the superscript $S_{n_f'}$.
The number of flavors $n_f'$ as the subscript of $S$ is the number of sea
quarks, while the $n_f$ as the subscript of $O$ and $w$ is the number of
valence quarks that characterizes the concrete form of operators.
For simplicity, we use vector and matrix notation as in the RHS of Eq.~\eqref{eq:HW}
by omitting the index $i$ of operators.
The superscript $T$ denotes the transposition of the vector or matrix.
The weak Hamiltonian is independent of $n_f$ in the sense that matrix elements
calculated in QCD between states which involve an energy scale $E$ do not
change when $n_f$ is increased above $n_f^{\mathrm eff}$, which is 
chosen so that quark flavors indexed by $n>n_f^{\mathrm eff}$ have
masses $m \gg E$.
%

If we calculate weak matrix elements with three-flavor
operators in $2+1$-flavor QCD ensembles,
we need the Wilson coefficients $w_3^{S_3}(\mu)$ in the three-flavor theory
to obtain the proper weak Hamiltonian
\begin{equation}
\left\langle \pi\pi|H_W|K\right\rangle
= {w_3^{S_3}(\mu)}^T\left\langle \pi\pi\Big|O_3^{S_3}(\mu)\Big|K\right\rangle_{2+1}.
\label{eq:ME}
\end{equation}
However, perturbative calculation of $w_3^{S_3}(\mu)$ requires
a matching from $w_4^{S_4}(\mu')$ that is performed below the charm threshold,
which induces a large systematic error ($\sim12\%$) \cite{Bai:2015nea}.
Therefore a non-perturbative matching in a non-perturbative scheme is desired.
The RI/(S)MOM scheme is not suitable since it cannot prevent mixing
with irrelevant operators allowed by a gauge-fixing and contact terms,
which may become more important at low scales.
A position-space scheme $X$ is a reasonable scheme to implement the
non-perturbative matching since it prevents significant mixing with
gauge-noninvariant operators and operators that are forbidden by the
equations of motion.

We consider the equality of two-point function
$\left\langle H_W(x)O_3^{X_{n_f}}(\mu;y)^\dag\right\rangle_{N_f}$
for 3 and 4 flavors:
\begin{equation}
{w_3^{S_3}(\mu_3)}^T
\left\langle O_3^{S_3}(\mu_3;x) {O_3^{X_3}(\mu;y)}^\dag\right\rangle_{2+1}
= 
{w_4^{S_4'}(\mu_4)}^T
\left\langle  O_4^{S_4'}(\mu_4;x){O_3^{X_4}(\mu;y)}^\dag\right\rangle_{2+1+1},
\end{equation}
which is valid at long distances $1/|x-y| \ll m_c$.
Then we obtain
\begin{align}
w_3^{S_3}(\mu_3)
= \left({Z_{O_3}^{S_3/\rm lat}(\mu_3,1/a)}^T\right)^{-1}
{G_{3\mathchar`-3}^{\rm lat[2+1]}(1/a;x-y)}
^{-1}
{Z_{O_3}^{X_3/\rm lat}(\mu,1/a)}^{-1}&
\notag\\* \times\,
Z_{O_3}^{X_4/\rm lat}(\mu,1/a)\,
G_{3\mathchar`-4}^{\rm lat[2+1+1]}(1/a;x-y)\,
{Z_{O_4}^{S_4'/\rm lat}(\mu_4,1/a)}^T&
w_4^{S_4'}(\mu_4),
\label{eq:W3}
\end{align}
where we define
\begin{equation}
G_{n_f\mathchar`-n_f'}^{\rm lat[N_f]}(1/a;x-y)
= \left\langle
O_{n_f}^{\rm lat}(1/a;x) {O_{n_f'}^{\rm lat}(1/a;y)}^\dag
\right\rangle_{N_f},
\end{equation}
and introduce renormalization matrices which satisfy
\begin{equation}
O^{S_{n_f'}}_{n_f}(\mu;x)
= Z^{S_{n_f'}/\rm lat}_{O_{n_f}}(\mu,1/a)O^{\rm lat}_{n_f}(1/a;x).
\end{equation}

If the sea charm quark is neglected, the relation becomes easier
\begin{align}
w_3^{S_3}(\mu_3)
\simeq \left({Z_{O_3}^{S_3/\rm lat}(\mu_3,1/a)}^T\right)^{-1}
&{G_{3\mathchar`-3}^{\rm lat[2+1]}(1/a;x-y)}^{-1}
\notag\\* &\times\,
G_{3\mathchar`-4}^{\rm lat[2+1]}(1/a;x-y)\,
{Z_{O_4}^{S_3'/\rm lat}(\mu_4,1/a)}^T
w_4^{S_4'}(\mu_4),
\label{eq:W3_Pquench}
\end{align}
so that renormalization matrices in the position-space scheme are not needed.
In addition, if the three-flavor operators are the same as the four-flavor
operators $O_3 = O_4$, {\it i.e.} if the valence charm quark cannot be
involved in the operators, the matching
between the three- and four-flavor Wilson coefficients is identical
as long as the sea charm quark is neglected.
On the other hand, if the charm quark is present in the 
four-flavor operators, as in the case of $K\to\pi\pi$ decays, the matching of
the Wilson coefficients is needed if the weak matrix elements are to be
calculated with the three-flavor operators.
Note that we actually need the lattice Wilson coefficients $w_3^{\rm lat}$,
which can be obtained from Eq.~\eqref{eq:W3_Pquench} if we simply drop
the multiplication by $(Z_{O_3}^{S_3/\rm lat})^{-1}$ from the RHS
of Eq.~\eqref{eq:W3_Pquench}, removing any reference to the
scheme $S_3$.

We will choose $S = S' = \rm RI/SMOM$, in which the Wilson coefficients
in the four-flavor theory can be calculated perturbatively.
To obtain the Wilson coefficients in the three-flavor theory, we need to
calculate the four-flavor renormalization matrix
$Z_{O_4}^{\rm RI/SMOM_3/lat}(\mu,1/a)$, the two-point Green's
functions of two three-flavor operators $G_{3\mathchar`-3}^{\rm lat[2+1]}(1/a;x-y)$
and those between a three-flavor operator and a four-flavor operator
$G_{3\mathchar`-4}^{\rm lat[2+1]}(1/a;x-y)$.
In the following sections, we fix $y=0$ for simplicity and
present our strategy to calculate the two-point 
functions with controlled discretization errors and the results from a
test calculation.

\section{Spherical average of two-point functions}
\label{sec:spheave}
The three-flavor Wilson coefficients calculated with the strategy proposed
in the previous section will depend on $x$, the relative distance between two
operators in the correlators.
In an ideal calculation with no discretization errors and sufficiently large
$m_c$ this $x$ dependence should be absent. 
In a practical calculation, $x$-dependence will arise because the charm
quark is insufficiently massive and from finite cut-off effects.
%
Although distance scale in $1/|x| \ll m_c$ is much longer than
recently used lattice spacings, discretization effects on correlators
at  $1/|x|\simeq 400$~MeV are more than 10\% depending on 
lattice spacing and much larger than statistical errors.
Thus, it is preferable to take the continuum limit to avoid such ambiguity.
However, in order to take the continuum limit, we need to calculate correlators
at a fixed physical distance for each lattice spacing, while
correlators on the lattice have values only at discrete points that depend
on lattice spacing.

We will apply the spherical averaging technique \cite{Tomii:2018zix}
to evaluate correlators at any physical distance as well as to reduce
discretization errors.
While correlators on the lattice violate $O(4)$ symmetry and depend
on lattice points in a complicated way, this technique enables us to 
obtain correlators that depend only on the distance $|x|$ as if they
have $O(4)$ symmetry.
There are two steps to evaluate sphere-averaged correlators using
lattice correlators $f_{a,n}$\footnote{
In Ref.~\cite{Tomii:2018zix}, we defined $f_{a,n}$ as a correlator
multiplied by $x^{2d}$ with the dimension $d$ of the operator
and divided by the same factor after taking the spherical average
to avoid strong $x$-dependence, which may induce a large discretization
error as a by-product of the spherical average.
}:
\begin{itemize}
\item {\bf Interpolation}\ \ \ \ 
In this step, we estimate the values of correlators at any physical location $x$. 
In the case of one dimension, it is easy to verify that the linear interpolation
\begin{equation}
\bar f_a(x) = \frac{(a(n+1)-x)f_{a,n}+(x-an)f_{a,n+1}}{a},
\label{eq:eq_wave1dim}
\end{equation}
cancels the $O(a^1)$ discretization error that arises from the Taylor
expansions of $f_{a,n}$ and $f_{a,n}$ around $x$.
In the case of four dimensions, the interpolation is modified to
\begin{equation}
\bar f_a(x) = a^{-4}\sum_{i,j,k,l=0}^1\Delta_{1,i}\Delta_{2,j}\Delta_{3,k}\Delta_{4,l}\,
f_{a,n+i\hat1+j\hat2+k\hat3+l\hat4},
\label{eq:quadrilin_interpolate}
\end{equation}
where $n_\mu = \lfloor x_\mu/a\rfloor$, $\hat\mu$ is the unit vector for
the $\mu$-direction and we define 
\begin{equation}
\Delta_{\mu,i} = |a(n_\mu+1-i)-x_\mu|.
\end{equation}
It is also easy to verify that this interpolation is free from $O(a^1)$ errors.
\item {\bf Average over spheres}\ \ \ \ 
While the interpolated correlators in four-dimensions designed above have values
at any physical location, they still violate rotational symmetry and depend on $x$
in a complicated way.
To obtain correlators as a continuous function of only the distance $|x|$, we
average them  over a four-dimensional sphere $U_{|x|}$ with the radius of $|x|$,
\begin{equation}
\hat f_a(|x|)
=\oint_{U_{|x|}}\td\sigma \bar f_a(x) \bigg/\oint_{U_{|x|}}\td\sigma.
\label{eq:sphe_ave4}
\end{equation}
\end{itemize}

\section{Exploratory calculation of two-point functions of four-quark operators}
\label{sec:4q2pt}

In this section, we show the result for a preliminary calculation of two-point
functions
\begin{equation}
G_{ij}(x) = \big\langle Q_i(x) Q_j(y)^\dag\big\rangle,
\end{equation}
of unrenormalized $\Delta S=1$ four-quark operators $Q_{i,j}$ in the three-flavor theory. 
In general, the calculation of these two-point functions requires all-to-all quark
propagators since there are diagrams that contain a quark loop at the sink point. 
Thus, there may be power divergence from loop diagrams, which needs
to be eliminated before renormalizing the operators.

\begin{figure}[tp]
\begin{center}
\subfigure{\mbox{\raisebox{1mm}{\includegraphics[width=110mm, bb=0 0 345 230]{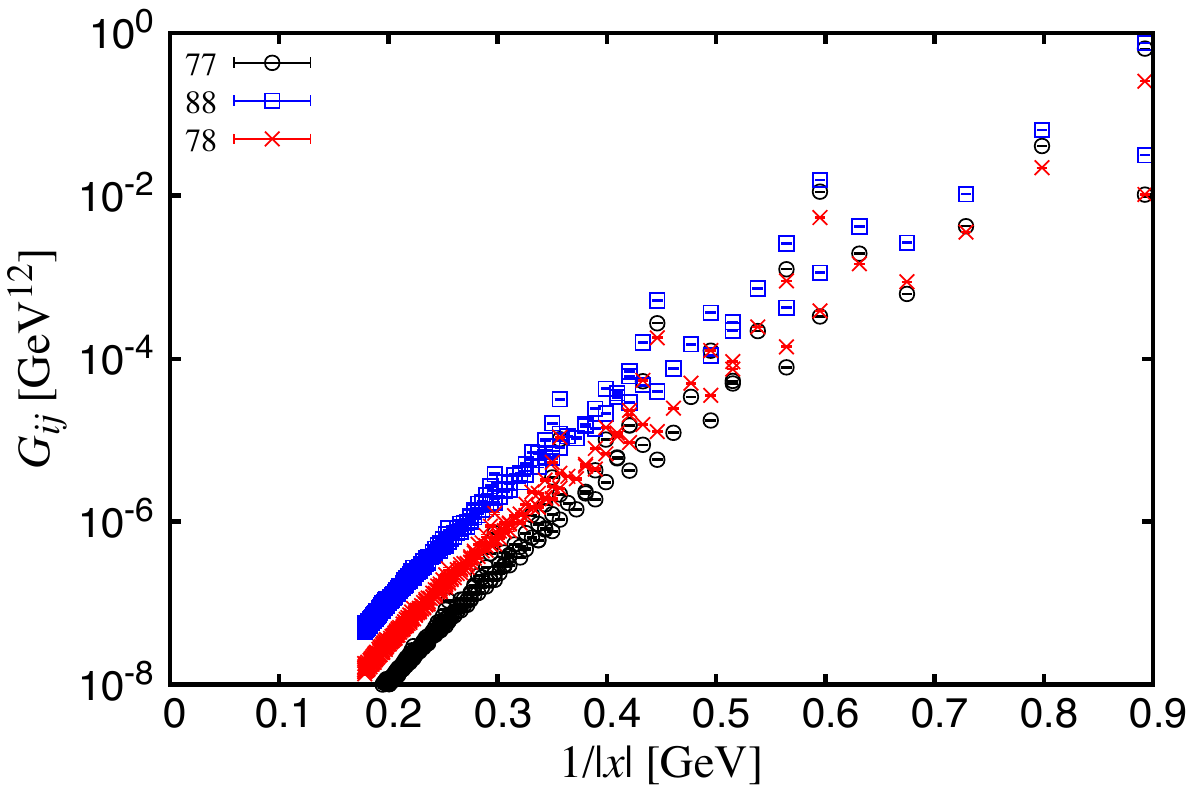}}}}
\subfigure{\mbox{\raisebox{1mm}{\includegraphics[width=110mm, bb=0 0 345 230]{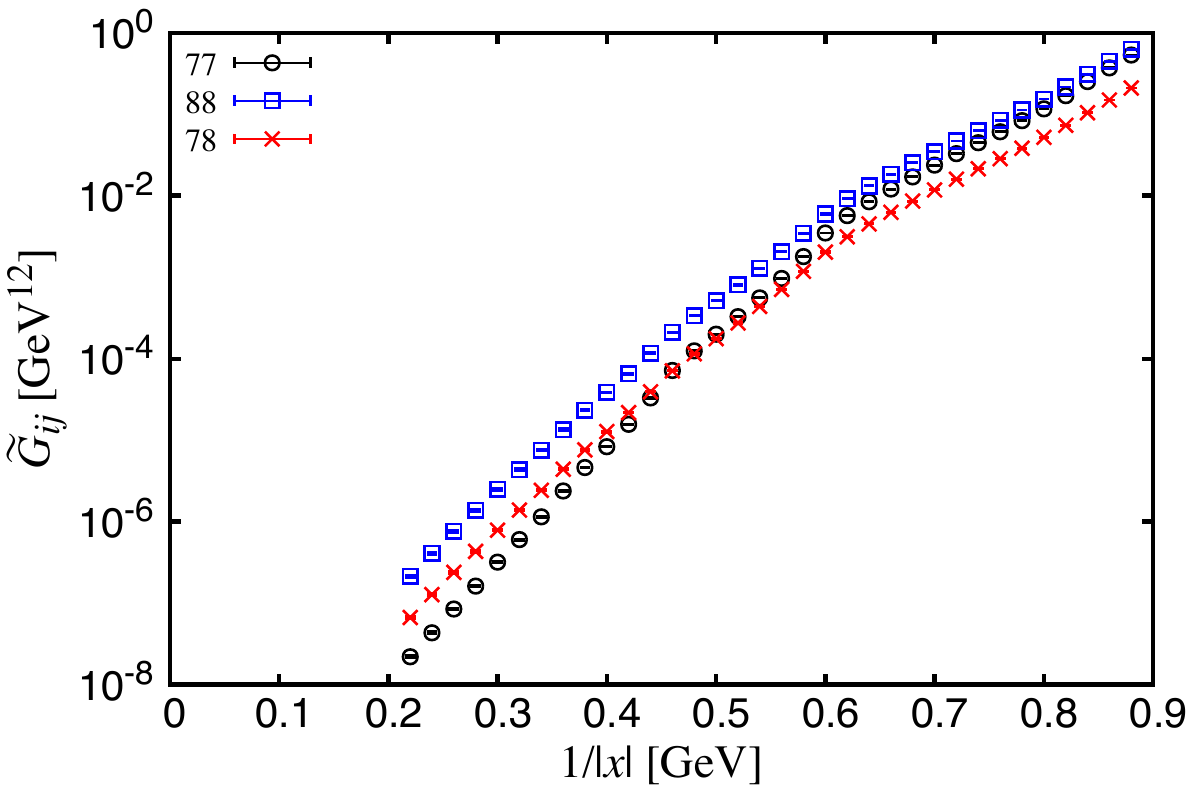}}}}
\caption{
Results for $G_{77}(x)$ (circles), $G_{88}(x)$ (squares) and $G_{78}(x)$ (crosses)
calculated on the coarsest lattice with $a^{-1} \simeq 1.79$ before (upper panel)
and after (lower panel) taking the spherical average.
}
\label{fig:8_8_coarsest}
\end{center}
\end{figure}

Among the $\Delta S = 1$ operators relevant for the $K\to\pi\pi$ matrix
elements, 
\begin{align}
Q_7 &= \frac{3}{2}
\overline s_\alpha\gamma_\mu(1-\gamma_5) d_\alpha
\sum_{q=u,d,s}e_q\overline q_\beta\gamma_\mu(1+\gamma_5) q_\beta,
\\
Q_8 &= \frac{3}{2}
\overline s_\alpha\gamma_\mu(1-\gamma_5) d_\beta
\sum_{q=u,d,s}e_q\overline q_\beta\gamma_\mu(1+\gamma_5) q_\alpha,
\end{align}
where $e_q$ is the electric charge of a quark $q$ the RHSs are summed
over the Lorentz index $\mu$ and the color indices $\alpha$ and $\beta$, 
enable us to investigate the simplest case of mixing correlator matrix
since only these two operators belong to the $(8_L,8_R)$ representation
of ${\rm SU}(3)_L\times{\rm SU}(3)_R$ symmetry
\cite{Blum:2001xb,Bernard1989ta}.
In this article, we show the result for the contribution of the fully-connected
diagrams in which all the quark propagators connect the source and sink
points and there is no power divergence.
(These are the only non-zero diagrams if we use the $I=3/2$ components
of $O_7$ and $O_8$.)

We use $2+1$-flavor domain-wall ensembles with three lattice cutoffs $a^{-1}$
ranging from 1.79~GeV to 3.15~GeV generated by the RBC and UKQCD
collaborations.
Pion masses are in the region from 300~MeV to 370~MeV.

\begin{figure}[tp]
\begin{center}
\subfigure{\mbox{\raisebox{1mm}{\includegraphics[width=75mm, bb=0 0 345 230]{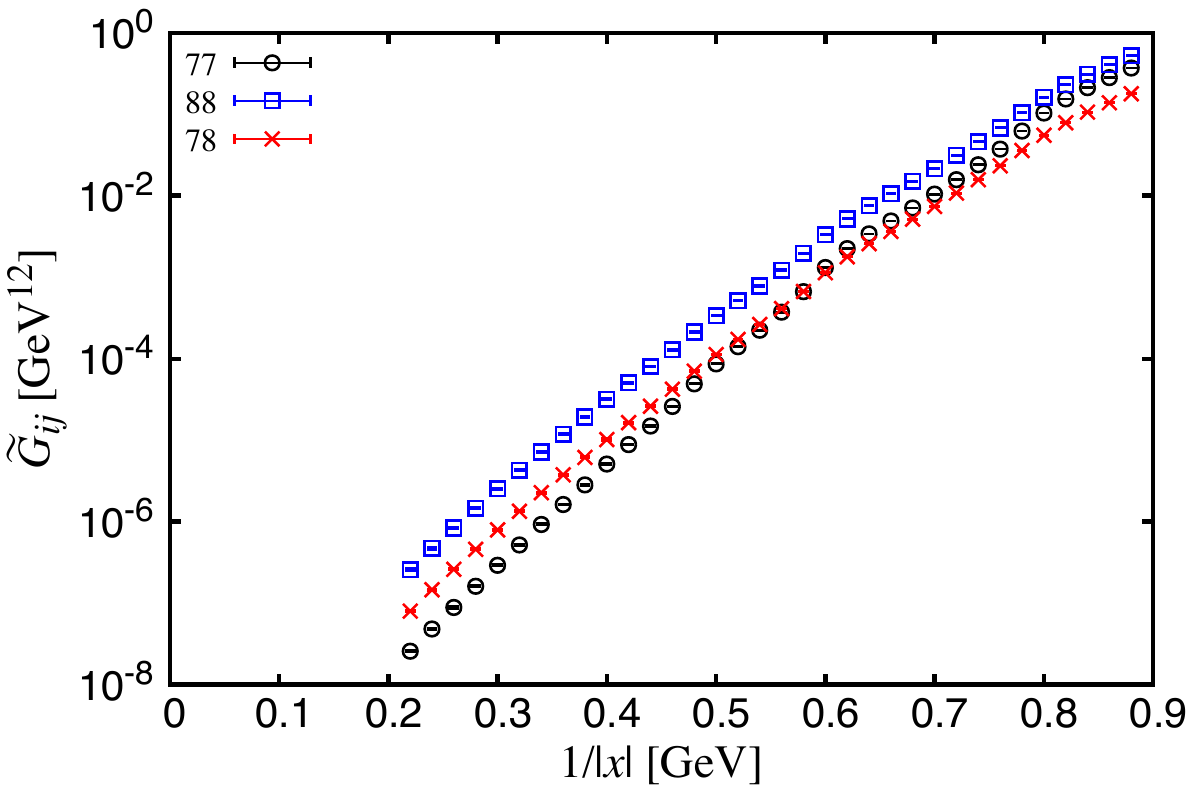}}}}
\subfigure{\mbox{\raisebox{1mm}{\includegraphics[width=75mm, bb=0 0 345 230]{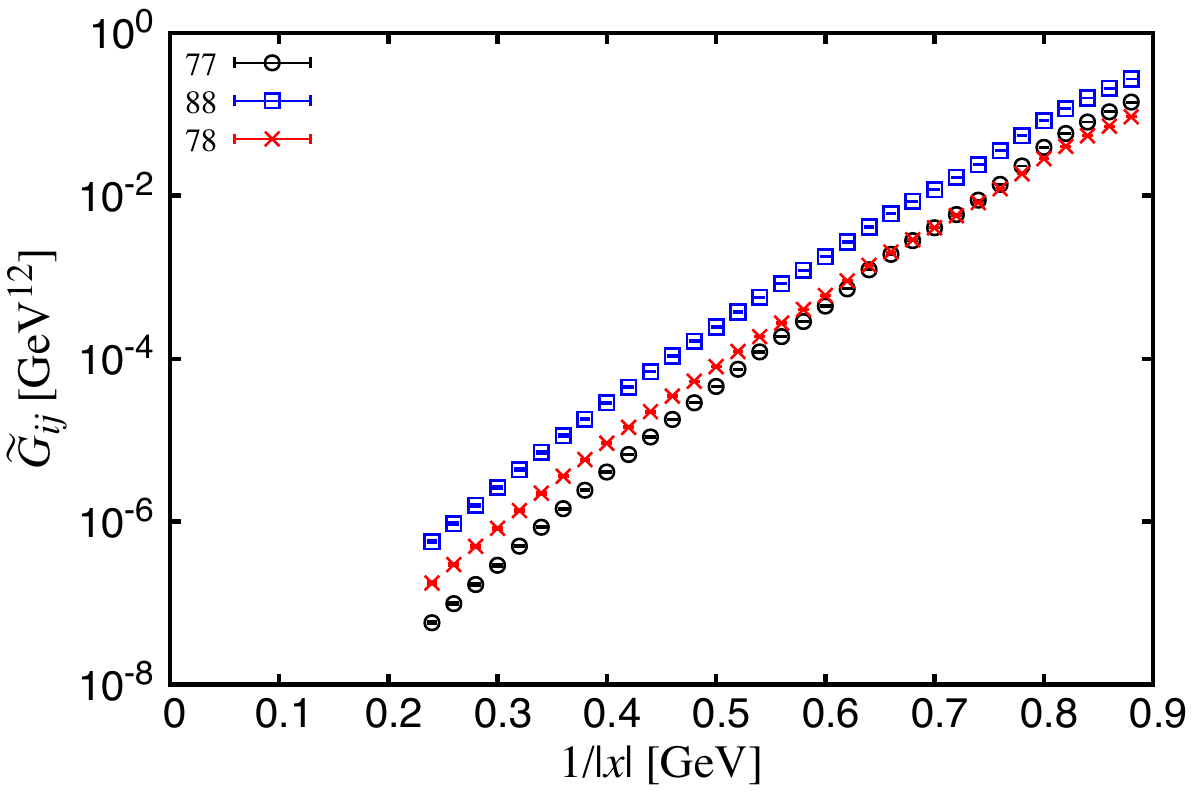}}}}
\caption{
Same as the lower panel of Figure~\ref{fig:8_8_coarsest} but the result on
finer lattices, $a^{-1} = 2.38$~GeV (left panel) and $a^{-1} = 3.15$~GeV (right panel).
}
\label{fig:8_8_finest}
\end{center}
\end{figure}

Figure~\ref{fig:8_8_coarsest} shows the results for two-point functions of
$(8_L,8_R)$ operators.
Since the correlator matrix is real symmetric, we take the average of the
78 and 87 elements, which is shown as the 78 element (crosses) in the figure.
The upper panel shows the results for $G_{ij}(x)$ before taking the
spherical average.
Here, we distinguish different lattice points that are not equivalent with respect 
to $90^\circ$ rotations or parity inversion in the four-dimensional hypercubic group. 
The results are averaged over sets of lattice points related by hypercubic
transformations.
The ambiguity due to the violation of $O(4)$ symmetry could amount to more than
$\times10$ at $|x| = 3a$ as numerical results at $|x| = 3a$ read
$G_{77}(0,0,0,3a) = 1.12(1)\times10^{-2}~{\rm GeV}^{12}$ and
$G_{77}(0,a,2a,2a) = 3.41(3)\times10^{-4}~{\rm GeV}^{12}$.
From the same observation, the ambiguity at $|x| = 6a$ is
about $\times 3$.
The lower panel shows the results for the spherical average
$\widetilde G_{ij}(|x|)$.
The discretization errors in the spherical average appear to be
much smaller than those in $G_{ij}(x)$.
Figure~\ref{fig:8_8_finest} show the results for the the spherical average
calculated on finer lattices, $a^{-1} = 2.38$~GeV (left panel) and
$a^{-1} = 3.15$~GeV (right panel).
As mentioned in the previous section, the spherical averaging technique
enables us to evaluate the values of correlators at any physical distance.
Therefore, the matching matrix between the Wilson coefficients in the
three- and four-flavor theories Eq.~\eqref{eq:W3} or \eqref{eq:W3_Pquench},
which is calculated from two-point functions, can easily be extrapolated
to the continuum limit at any physical distance $|x|$.

\section{Summary}
\label{sec:summary}

We formulate a non-perturbative strategy to match the three- and four-flavor
Wilson coefficients of $\Delta S = 1$ four-quark operators.
We propose to use two-point Green's functions of four-quark
operators and their spherical average to take the
continuum limit of the matching matrix.
As Eq.~\eqref{eq:W3_Pquench} indicates, we also needs the renormalization
matrix of the four-flavor operators in a scheme
in which perturbative calculation is available.
The four-flavor operators will be calculated as well as the two-point functions
in near future.

\end{document}